\documentclass[aps,prl,prabib,twocolumn,
showpacs,preprintnumbers,amsmath,amssymb,floatfix,aps]{revtex4}

\usepackage{graphicx}

\newcommand{\up}{\uparrow}
\newcommand{\down}{\downarrow}
\newcommand{\uden}{n_\uparrow}
\newcommand{\dden}{n_\downarrow}
\newcommand{\eup}{E_{F \uparrow}}
\newcommand{\br}{{\bf r}}
\newcommand{\bp}{{\bf p}}

\begin{document}

\title{Normal state of a polarized Fermi gas at unitarity}
\author{C. Lobo, A. Recati, S. Giorgini, and S. Stringari}
\affiliation{Dipartimento di Fisica, Universit\`a di Trento
and CNR-INFM BEC Center, I-38050 Povo, Trento, Italy}

\begin{abstract} 
We study the Fermi gas at unitarity and at $T=0$ by assuming that, at high
polarizations, it is a normal Fermi liquid composed of weakly
interacting quasiparticles associated with the minority spin
atoms. With a quantum Monte Carlo approach we calculate their
effective mass and binding energy, as well as the full equation of
state of the normal phase as a function of the concentration
$x=n_\down/n_\up$ of minority atoms. We predict a first order phase
transition from normal to superfluid at $x_c=0.44$ corresponding, in
the presence of harmonic trapping, to a critical polarization
$P_c=(N_\uparrow-N_\downarrow)/ (N_\uparrow+N_\downarrow)=77\%$.  We
calculate the radii and the density profiles in the trap and predict
that the frequency of the spin dipole mode will be increased by a
factor of 1.23 due to interactions.
\end{abstract}

\maketitle 

Recent experiments on degenerate gases of $^6$Li with a mixture of two
hyperfine species have explored the physics of Fermi gases
\cite{experiment1,huletexp,experiment2} and have led to a number of
theoretical analyses \cite{theory1,yip,mueller,theory2}. One of the
major experimental observations has been the occurrence of phase
separation if the mixture contains more atoms of one species than of
the other, i.e., if the gas is polarized. Some of the experiments
suggest that in the unitary limit of strong interactions there are
three phases: an unpolarized superfluid phase, a mixed phase which
exhibits a partial polarization and a fully polarized gas. We now have
a good understanding of the superfluid phase which has been the subject
of numerous theoretical and experimental studies while the fully
polarized phase is an ideal Fermi gas since atoms in the same spin
state do not interact with each other. However, for intermediate
polarizations, when both species are present, the nature of the mixed
phase is not understood.

Here we study the mixed phase in the unitary limit by
adopting an approach inspired by the theory of dilute solutions of
$^3$He in $^4$He \cite{bardeen}. We will assume that the majority
species ($\uparrow$) forms a background experienced by the minority
species ($\downarrow$) and that the latter behaves as a gas of weakly
interacting fermionic quasiparticles even though the $\up-\down$
atomic interaction is very strong, being characterized by an infinite
scattering length. In other words, we will assume that the system is a normal
Fermi liquid, which will allow us to characterize the energy of the gas
in terms of a few parameters and, by calculating these with a quantum Monte
Carlo approach, allow us to make various predictions of experimental
relevance.

We begin by writing the expression for the energy $E$ of a homogeneous
system in the limit of very dilute mixtures and at zero
temperature. The concentration of $\down$ atoms is given by the ratio
of the densities $x=\dden/\uden$ and we will take it to be small. If
only $\up$ atoms are present then the energy is that of an ideal Fermi
gas $E(x=0)=3/5 \eup N_\up$, where $N_\up$ is the total number of
$\up$ atoms and $\eup=\hbar^2/2m (6 \pi^2 \uden)^{2/3}$ is the ideal
gas Fermi energy.  When we add a $\down$ atom with a momentum ${\bf
p}$ ($|\bp| \ll p_{F \up}$), we shall {\em assume} that the change in
$E$ is given by
\begin{equation}
\delta E=\frac{p^2}{2m^*}-\frac{3}{5} \eup A. \label{eq:deltaE}
\end{equation}
In other words, the $\down$ atom in the $\up$ gas behaves as a
quasiparticle with a quadratic dispersion and an effective mass
$m^*$. In addition, there is a ``binding'' energy $-3/5 \eup A$ of the
$\down$ atom to the Fermi gas of $\up$ atoms. This binding energy must
be proportional to $\eup$ since there is no other energy scale in the
unitary limit and we have used the factor $3/5$ for later
convenience. We shall further assume that this quasiparticle is a
fermion \cite{statistics}.

When we add more $\down$ atoms, creating a small finite density
$\dden$, they will form a degenerate gas of quasiparticles at zero
temperature occupying all the states with momentum up to the Fermi momentum
$p_{F\down}=\hbar(6 \pi^2 \dden)^{1/3}$. The energy of the system can
then be written in a useful form in terms of the concentration $x$ as:
\begin{equation}
\frac{E(x)}{N_\up}=\frac{3}{5}\eup \left(1 - A
x+\frac{m}{m^*}x^{5/3} \right). \label{eq:energyx}
\end{equation}
Eq.(\ref{eq:energyx}) is valid for small values of the concentration
$x$, i.e. when interactions between $\down$ quasiparticles as well as further
renormalization effects of the parameters can be neglected.

\begin{figure}[b]
\begin{center}
\includegraphics[width=6.5cm]{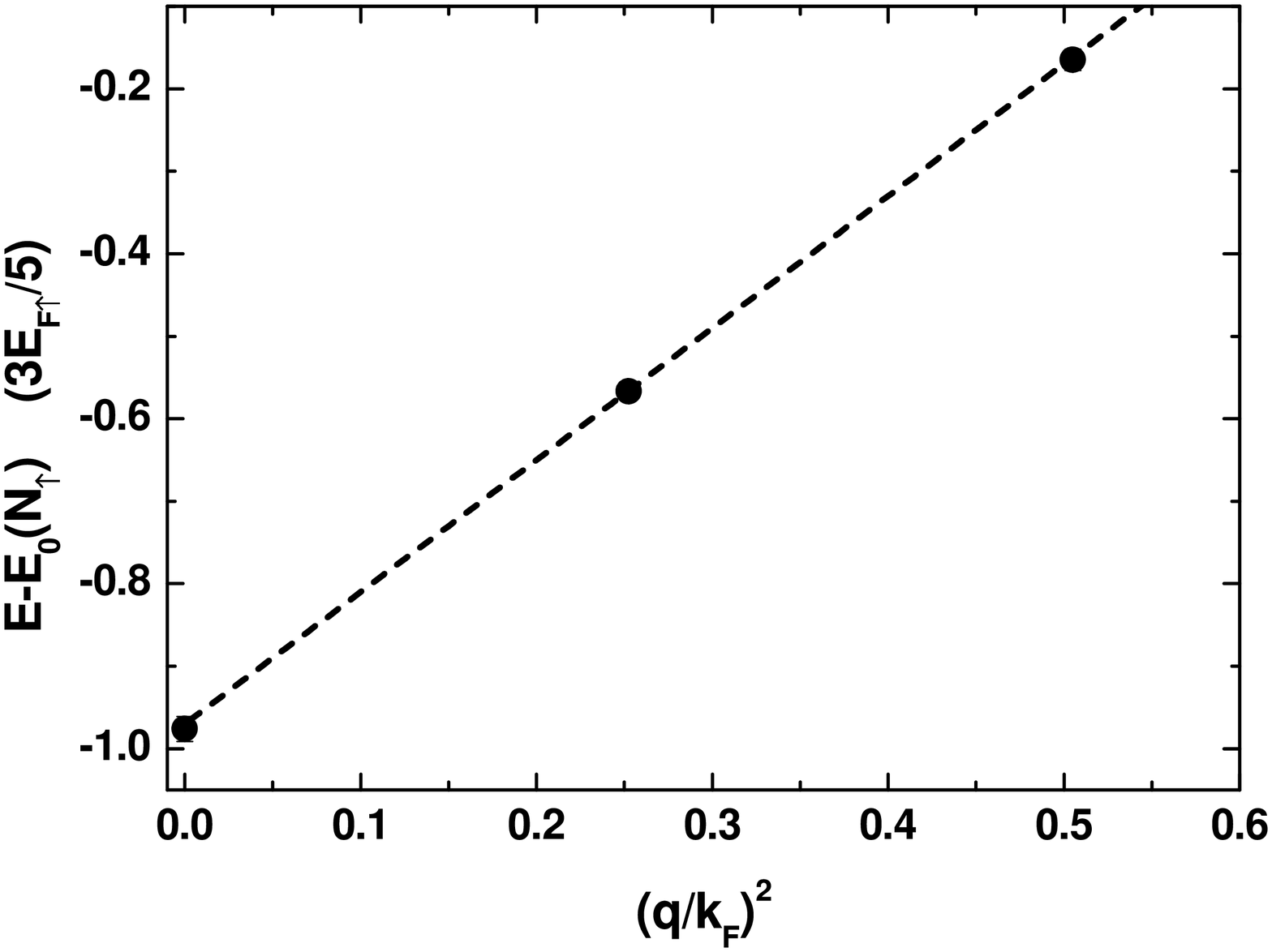}
\caption{Excitation spectrum of a $\downarrow$ atom in a Fermi sea
of $\uparrow$ atoms. The dashed line is a linear best fit to the
FN-DMC results from which we extract the values $A$ and $m^*$.}
\label{fig1}
\end{center}
\end{figure}

In the following we will calculate $A$ and $m^*$ using a fixed-node
diffusion Monte Carlo (FN-DMC) approach, already employed in earlier
studies~\cite{QMC1,QMC2}.  We use the same attractive square-well
potential to model the interactions between $\downarrow$ and
$\uparrow$ atoms: $V(r)=-V_0$ for $r<R_0$ and $V(r)=0$ otherwise. The
short range $R_0$ is chosen as $2n_\uparrow R_0^3=10^{-6}$. The depth
$V_0$ is fixed by the unitarity condition $|a|=\infty$ for the
$s$-wave scattering length $a$ and corresponds to the threshold for
the first two-body bound state in the well:
$V_0=\pi^2\hbar^2/(4mR_0^2)$. For a single $\downarrow$ atom in a
homogeneous Fermi sea of $\uparrow$ atoms the trial wave function
$\psi_T$, which determines the nodal surface used as an {\it ansatz}
in the FN-DMC calculation, is chosen to be of the form \cite{boronat}
\begin{equation}
\psi_T^{\bf q}({\bf r}_{\downarrow},{\bf r}_{1\up},...,{\bf
r}_{N_\uparrow})=\exp{(i{\bf q}\cdot r_{\downarrow})}
\prod_{i=1}^{N_\uparrow}f(r_{i\downarrow})D_{\uparrow}(N_\uparrow)\;,
\label{tf1}
\end{equation}
where $r_\downarrow$ denotes the position of the $\downarrow$ atom and
$r_{i\downarrow}= |{\bf r}_\downarrow-{\bf r}_{i\up}|$. In this
equation the plane wave $\exp{(i{\bf q}\cdot r_{\downarrow})}$
corresponds to the impurity travelling through the medium with
momentum $\hbar{\bf q}=(n_x,n_y,n_z)2\pi\hbar/L$, where $L$ is the
lenght of the cubic box and the $n_i$ are integers describing the
momentum in each coordinate. Furthermore, $D_\uparrow(N_\uparrow)$ is
the Slater determinant of plane waves describing the Fermi sea of the
$N_\uparrow$ atoms and the Jastrow term $f(r)$ accounts for
correlations between the impurity and the Fermi sea. The correlation
function $f(r)$ is chosen as in Ref.~\cite{QMC1}.  We consider a
system of $N_\uparrow=33$ atoms and periodic boundary conditions. From
the energy $E$ of the system with $N_\uparrow +1$ atoms we subtract
the exact energy $E_0(N_\uparrow)$ of the Fermi sea. The result is
shown in Fig.~\ref{fig1} as a function of $(q/k_{F\uparrow})^2$. From
a linear best fit we obtain the following values: $A=0.97(2)$ and
$m^*/m=1.04(3)$ which is consistent with the general inequality
$m^*/m \geq 1$. We have checked that finite-size corrections
associated with the number $N_\uparrow$ of atoms in the Fermi sea are
below the reported statistical error.  It is worth noticing that the
binding energy of a single $\downarrow$ atom almost coincides with the
average energy of an atom in the Fermi sea. A similar coincidence was
found for the superfluid gap $\Delta$ at unitarity~\cite{Carlson}.

A relevant question is to understand whether the equation of state
(\ref{eq:energyx}) is adequate to describe regimes of large values of
$x$ where interaction between quasiparticles and other effects might become
important. To answer this question we have carried out a FN-DMC
calculation of the equation of state at finite concentrations
$x=N_\downarrow/N_\uparrow$ using the trial wave
function
\begin{equation}
\psi_T({\bf r}_{1^\prime},...,{\bf r}_{N_\downarrow},{\bf
r}_1,...,{\bf r}_{N_\uparrow})=
\prod_{i,i^\prime}f(r_{ii^\prime})D_{\downarrow}(N_\downarrow)D_{\uparrow}
(N_\uparrow)\;,
\label{tf2}
\end{equation}
where $i$ and $i^\prime$ label, respectively, $\uparrow$ and
$\downarrow$ atoms. The nodal surface of the wave function $\psi_T$ is
determined by the product of Slater determinants
$D_\uparrow(N_\uparrow)D_\downarrow(N_\downarrow)$ and coincides with
the nodal surface of a two-component ideal Fermi gas. As a
consequence, the wave function in Eq. (\ref{tf2}) is incompatible with
off-diagonal long-range order (ODLRO) and describes a normal Fermi
gas.  In contrast, the BCS-type wave function used in
Refs.~\cite{QMC1,QMC2} is compatible with ODLRO and describes a
superfluid state. A direct comparison between the groundstate energy
of the normal and superfluid states can be carried out for equal
numbers of $\uparrow$ and $\downarrow$ atoms,
$N_\downarrow=N_\uparrow$, with the result $E_{\rm
SF}/(3/5\eup N_\up)= 0.84(2)$ and $E_{\rm
N}/(3/5\eup N_\up)=1.12(2)$ showing the instability of the
normal state for $x=1$ (see Fig.~\ref{fig2}).

The results for the equation of state of the normal Fermi gas are
shown in Fig.~\ref{fig2}. To
reduce finite-size effects we have considered closed-shell
configurations $N_\downarrow$=7,19,27,33 with $N_\uparrow$=27,33. In
Fig.~\ref{fig2} we also show the prediction of Eq.(\ref{eq:energyx})
based on noninteracting quasiparticles (dashed line). For small
values of $x$ we find very good agreement, but for larger
concentrations effects of interactions between quasiparticles start to
be important and deviations from Eq.(\ref{eq:energyx}) become
visible. The solid line is obtained from a polynomial best fit to the
FN-DMC results.

From the equation of state of the mixed phase we can determine the
transition between the fully polarized and the mixed phases as well as
the transition between the mixed and the unpolarized superfluid phases
\cite{psf}. The equilibrium condition is obtained by imposing that the
chemical potential and the pressure be the same in the two phases.  It
is useful to express the results in terms of the chemical
potential $\mu=(\mu_\up+\mu_\down)/2$ and of the effective magnetic
field $h=(\mu_\up-\mu_\down)/2$.  Here $\mu_{\up \down}=\partial
E/\partial N_{\up \down}$ is the chemical potential of each spin
species.  The transition between the fully polarized and the mixed
phase is second order and takes place at $x=0$ where we find
$\mu_\down/\mu_\up=-3/5 A$, corresponding to
$h/\mu=3.78$.  The transition between the mixed
and the unpolarized superfluid phases is instead first order and is
simply obtained via the standard Maxwell construction considering the
tangent to the equation of state of Fig.~\ref{fig2} crossing the
superfluid point $E_{\rm SF}/(3/5\eup N_\uparrow)=0.84$ at $x$=1. We
obtain the critical value $x_c=0.44$ at the transition.  For smaller
values of $x$ the system remains in the normal state, while above the
critical concentration $x_c$ the system will begin nucleating the
superfluid and phase separate into those two states.  The phase
transition is characterized by the value
$\mu_\downarrow/\mu_\uparrow=0.017$ corresponding to $h/\mu=0.96$
\cite{Chevy}. Note that at the critical value $x_c=0.44$ the
difference between the best fit and the prediction of
Eq.({\ref{eq:energyx}) is quite small so that this latter energy
functional describes well the whole normal phase. This supports the
idea that the normal state is a gas of weakly
interacting quasiparticles.  Eq.({\ref{eq:energyx}) would actually
predict the value $x_c=0.50$ for the critical concentration, which
corresponds to $h/\mu=0.93$.
\begin{figure}[b]
\begin{center}
\includegraphics[width=6.5cm]{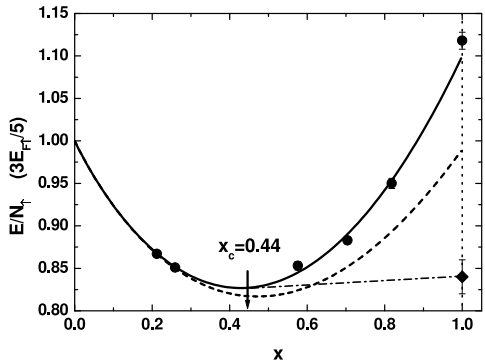}
\caption{Equation of state of a normal Fermi gas as a function of the
concentration $x$ (circles). The solid line is a polynomial best fit to the
FN-DMC results. The dashed line corresponds to
Eq.(\ref{eq:energyx}). The dot-dashed line is the coexistence line
between the normal and the unpolarized
superfluid states and the arrow indicates the critical concentration
$x_c$ above which the system phase separates. For $x=1$, both the
energy of the normal and of the superfluid (diamond) states are shown.}
\label{fig2}
\end{center}
\end{figure}

It is useful to compare the above results with BCS theory which also
predicts 3 phases at unitarity \cite{yip,mueller}: a superfluid with
energy $E_{\rm BCS}/(3/5 \eup N_\up)=1.18$, a mixed state which is
a noninteracting partially polarized gas and a fully polarized
gas. The mixed state energy is simply the kinetic energy and is an
increasing function of $x$. The tangent to the curve is at $x_c=0.04$,
corresponding to $\mu_\up/\mu_\down=0.1$, thereby leading to a much
reduced normal region with respect to the predictions of our
FN-DMC calculation.

Now we turn to the trapped case. While we will discuss the situation
only at zero temperature we should note that temperature can have an
important effect on the density profile of the $\down$ atoms. In
experiments we usually have the condition $k_B T \ll \eup$. But since
$\dden \ll \uden$, we might be in a situation where $k_B T \gtrsim
E_{F \down}$ and would therefore need to use the appropriate thermal
distribution. With this caveat we turn to the inhomogeneous situation
where we shall use the local density approximation (LDA)
\cite{hulet}. In a harmonic trap with potential $V(\br)=\sum_i m
\omega^2_i r_i^2/2$, the local chemical potentials become $\mu_{\up
\down}(\br)= \mu_{\up \down}^0 -V(\br)$. For small concentrations (in
particular $x\ll x_c$), where only the normal state is present and
where we can neglect the change in $\mu_\up$ due to the attraction of
the $\down$ atoms, $\uden$ is the Thomas-Fermi density of an ideal gas
whereas $\dden$ is a Thomas-Fermi profile with a modified harmonic
potential $V \rightarrow V \times (1+3/5 A)$. The potential seen
by the $\down$ atoms is more confining due to the attraction to
the $\up$ atoms. This also has consequences for the collective modes
of the system: it leads us to predict that the spin dipole mode - the
mode where the $\down$ atoms oscillate as a whole in the midst of the
$\up$ atom cloud - will have a frequency given by
\begin{equation}
\omega_i^{\rm dip}= \omega_i \sqrt{(1+3/5 A)\frac{m}{m^*}} \simeq 1.23
\omega_i. \label{eq:frequency}
\end{equation}
So, a direct measurement of the oscillation frequency of the $\down$
atoms in a dilute mixture with $x \ll 1$ would provide a useful test
of this Fermi liquid theory and in particular of the numerical
estimate of the parameters $A$ and $m^*$ \cite{damping}.
\begin{figure}
\begin{center}
\includegraphics[width=7cm]{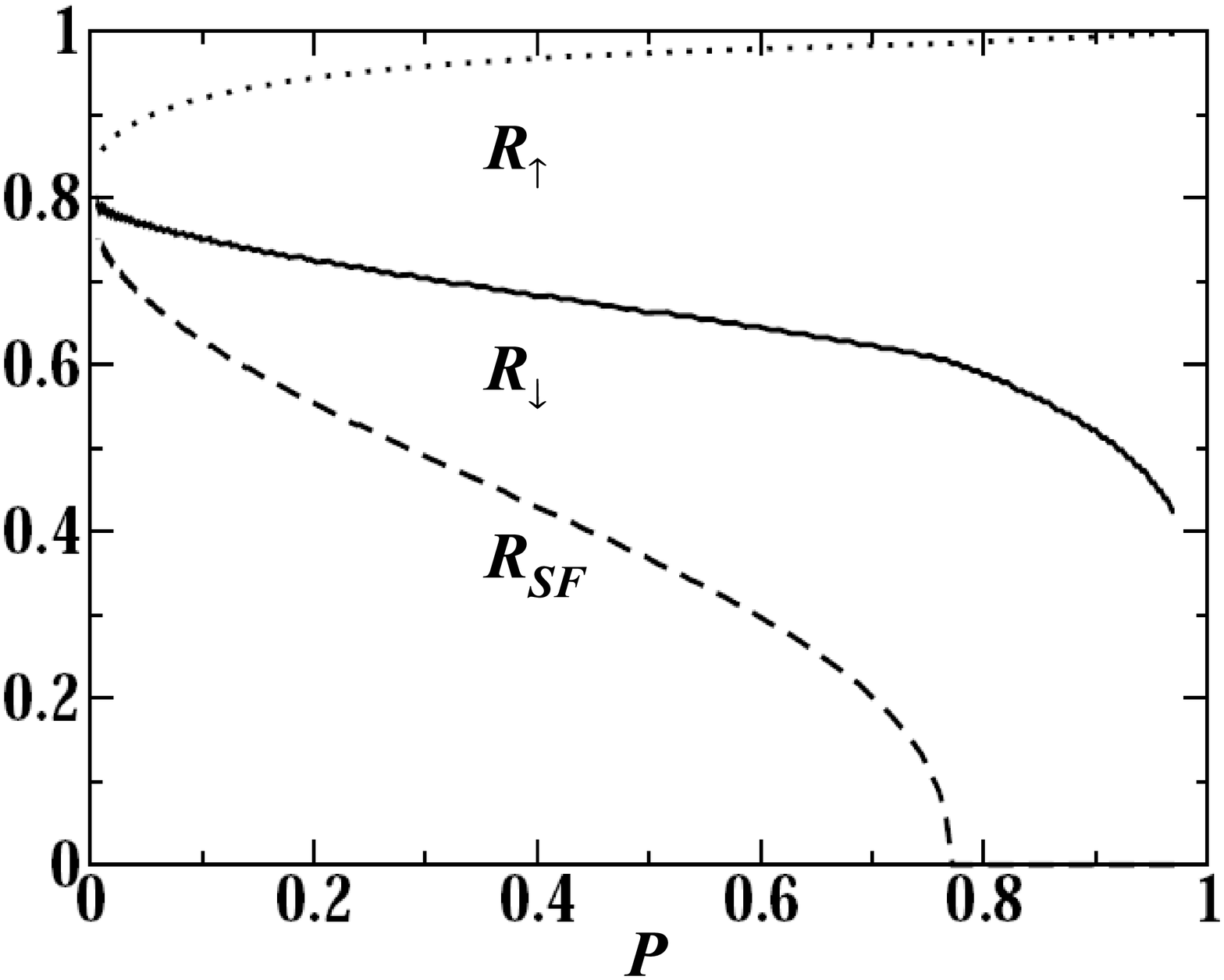}
\caption{Radii of the three phases in the trap in units of the radius
$R_\up^0=a_{\rm ho}(48 N_\up)^{1/6}$ of a noninteracting fully
polarized gas, where $a_{\rm ho}$ is the harmonic oscillator length.}
\label{radii}
\end{center}
\end{figure}
For larger concentrations the system will exhibit a central superfluid
core with chemical potential $\mu_{\rm SF}=(\mu_\up+\mu_\down)/2$,
whose radius in a spherical trap is given by
\begin{equation}
R_{SF}^2=R_\up^2\frac{\mu_\down^0/\mu_\up^0-\mu_\down/\mu_\up}
{1+\mu_\down/\mu_\up},
\end{equation}
where $R_\up$ is the radius of the $\up$ component and
$\mu_\down/\mu_\up$ is calculated at the transition point.  In
Fig.~\ref{radii}, we plot the radius $R_{SF}$ of the superfluid
component together with the radii of the minority and majority
components, $R_\up$ and $R_\down$ respectively, in units of
$R_\up^0=a_{\rm ho}(48 N_\up)^{1/6}$ as a function of the polarization
$P=(N_\up-N_\down)/(N_\up+N_\down)$ of the sample. We predict that the
superfluid phase disappears at $P_c=0.77$ in good agreement with the
experimental findings of \cite{experiment2}. Notice that as $P\to 1$,
$R_\down \sim R_\up^0 \left[ (1+3/5 A)
m^*/m\right]^{-1/4}\left[(1-P)/(1+P)\right]^{1/6}\to 0$ while $R_\up$
approaches the noninteracting value. In the opposite $P\to 0$ limit
the radii converge to the known value $(\mu/\eup)^{1/4}\simeq 0.80$.
It is worth noticing that the BCS approach would predict the value
$P_c>0.99$ at unitarity \cite{yip,mueller} pointing out the dramatic
role played by the binding energy in the mixed state.

At the transition between the superfluid and the mixed normal phase
the density of the two spin species exhibits a discontinuity,
revealing the first order nature of the transition.  The densities
$n_\up$ and $n_\down$ jump from the superfluid value $n_{SF}$ to the
values $n_\up\simeq 1.01 n_{SF}$ and $n_\down=x_cn_\up\simeq
0.44n_{SF}$ respectively, as one enters the normal phase. The
discontinuity is an artifact of LDA and the inclusion of surface
energy effects could    In
Fig.~\ref{fig3} we plot $\uden$ and $\dden$ as well as the difference
$\uden-\dden$ as a function of position in a spherical trap when
$P=P_c$. These results, based on LDA, apply also to anisotropic traps
through a simple scaling transformation. We find that both the total
density $\uden+\dden$ and the density difference $\uden-\dden$
increase monotonically towards the center \cite{discrepancy}. If
$P<P_c$ then $\uden=\dden$ in the central superfluid region.
\begin{figure}
\begin{center}
\includegraphics[width=7cm]{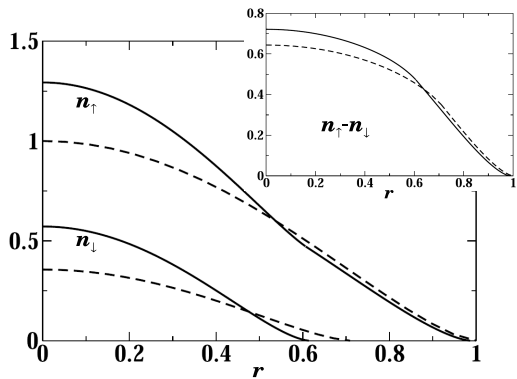}
\caption{Density profiles (solid lines) for the two spin components in
a spherical harmonic trap at the critical total polarization
$P_c=0.77$. The density profiles (dashed lines) for a noninteracting
Fermi gas with the same polarization are also shown. The inset shows
the density difference profile. Densities are given in units of the
central density of the noninteracting gas. The radial coordinate is
given in units of $R^0_\down$ (see Fig.~\ref{fig3}).}
\label{fig3}
\end{center}
\end{figure}

In conclusion, we study the polarized Fermi gas at unitarity as a
normal Fermi liquid composed of weakly interacting quasiparticles
associated with the minority atoms. These have a quadratic dispersion
and, in the $x \rightarrow 0$ limit, have an effective mass $m^*\simeq
1.04(3) m$ with a binding energy $-3/5 \eup \times 0.97(2)$,
calculated with a FN-DMC approach.  We derive an energy functional
Eq.(\ref{eq:energyx}) with those parameters, assuming noninteracting
quasiparticles and, using FN-DMC calculations at higher values of $x$,
show that corrections to the energy functional are small, even for
relatively high concentrations. Assuming that no polarized superfluid
phases exist, we predict a normal/superfluid first order phase
transition at a critical value $x_c=0.44$, corresponding, in the
presence of harmonic trapping, to a critical total polarization
$P_c=0.77$.  We also predict for small concentrations an increase by a
factor of 1.23 of the frequency of the spin dipole mode with respect
to the noninteracting value. A further application of the equation of
state of the mixed phase could be, for example, the calculation of the dipole
polarizability of the trapped Fermi gas \cite{dipole}.

We acknowledge useful discussions with I. Carusotto, A. J. Leggett,
L.P. Pitaevskii and N. Prokof'ev.  We also acknowledge support by the
Ministero dell'Istruzione, dell'Universit\`a e della Ricerca (MIUR).


\begin{thebibliography}{99}
\bibitem{statistics} Adding a $\down$ atom to a gas of $\up$ atoms in
the BEC limit would lead instead to the formation of a molecule. We
would then have a mixture of $\up$ atoms and a Bose-Einstein
condensate of molecules.
\bibitem{psf} We do not take into account the possibility of a polarized
superfluid which would modify the possible transitions.
\bibitem{Chevy} Upper and lower bounds for $\mu_\down/\mu_\up$ at the
the fully polarized/mixed and mixed/superfluid transitions (which are
compatible with our values) were recently given in Ref.\cite{theory2}.
\bibitem{damping} One concern is the possibility that this mode be
overdamped due to the nonsuperfluid nature of the system.  However, at
$k_BT \ll \eup$ we expect that the damping of the oscillation will be
sufficiently small.
\bibitem{discrepancy} A preliminary comparison with the
profiles of \cite{experiment2} reveals a qualitative discrepancy: it
is seen experimentally that the density difference has a minimum in
the center also at high polarizations. This might be a
signature of a polarized superfluid phase.
\end{thebibliography}
\end{document}